\documentclass[conference]{IEEEtran}
\usepackage{cite}
\usepackage{soul} 
\usepackage{url}
\usepackage{amsmath,amssymb,amsfonts}
\usepackage{array}
\usepackage{blkarray} 
\usepackage{makecell} 
\usepackage{algorithmic}
\usepackage{graphicx}
\usepackage{textcomp}
\usepackage{xcolor}
\usepackage{booktabs}
\def\BibTeX{{\rm B\kern-.05em{\sc i\kern-.025em b}\kern-.08em
    T\kern-.1667em\lower.7ex\hbox{E}\kern-.125emX}}

\ifCLASSOPTIONcompsoc
 \usepackage[caption=false,font=normalsize,labelfont=sf,textfont=sf]{subfig}
\else
 \usepackage[caption=false,font=footnotesize]{subfig}
\fi

\begin{document}

\title{Error-Mitigated Quantum Random Access Memory}

\author{\IEEEauthorblockN{Wenbo Shi\IEEEauthorrefmark{1}\thanks{Corresponding author: Wenbo Shi (email: wenbo.shi@unsw.edu.au).},
Neel Kanth Kundu\IEEEauthorrefmark{2}\IEEEauthorrefmark{3},
Matthew R. McKay\IEEEauthorrefmark{3}, and
Robert Malaney\IEEEauthorrefmark{1}}\\
\IEEEauthorblockA{\IEEEauthorrefmark{1}School of Electrical Engineering and Telecommunications,
\textit{University of New South Wales}, Sydney, NSW, Australia.}
\IEEEauthorblockA{\IEEEauthorrefmark{2}Centre for Applied Research in Electronics, \textit{Indian Institute of Technology Delhi}, New Delhi, India.}
\IEEEauthorblockA{\IEEEauthorrefmark{3}Department of Electrical and Electronic Engineering, \textit{University of Melbourne, Melbourne}, Victoria, Australia.}}


\maketitle

\begin{abstract}

As an alternative to quantum error correction, quantum error mitigation methods, including Zero-Noise Extrapolation (ZNE), have been proposed to alleviate run-time errors in current noisy quantum devices. In this work, we propose a modified version of ZNE that provides for a significant performance enhancement on current noisy devices. Our modified ZNE method extrapolates to zero-noise data by evaluating groups of noisy data obtained from noise-scaled circuits and selecting extrapolation functions for each group with the assistance of estimated noisy simulation results. To quantify enhancement in a real-world quantum application, we embed our modified ZNE in Quantum Random Access Memory (QRAM) - a memory system important for future quantum networks and computers. Our new ZNE-enhanced QRAM designs are experimentally implemented on a 27-qubit noisy superconducting quantum device, the results of which demonstrate QRAM fidelity can be improved significantly relative to traditional ZNE usage. Our results demonstrate the critical role the extrapolation function plays in ZNE - judicious choice of that function on a per-measurement basis can make the difference between a quantum application being functional or non-functional.

\end{abstract}


\section{Introduction}

Current quantum devices are widely regarded as Noisy Intermediate-Scale Quantum (NISQ) devices - quantum-enabled systems that do not have low enough intrinsic errors to implement quantum error correction~\cite{Preskill2018NISQ}.
To reach a quantum advantage using current NISQ devices, growing attention is being focused on quantum error mitigation methods, schemes that attempt to reduce the presence of intrinsic device errors~\cite{Czarnik2021errormitigation, Qin_2022}. In general,
quantum error mitigation generates a number of ancillary quantum circuits and applies classical post-processing to the measurement outcomes of the circuits in an attempt to deduce zero-noise results.
Such mitigation is known to be partially effective in reducing intrinsic errors within many current NISQ devices, especially when the intrinsic errors mainly consist of quantum gate errors, measurement errors, decoherence errors, and/or cross-talk errors~\cite{Wood2020ErrorType, Wilson2020ErrorType}.

The most common methods discussed for quantum error mitigation include Zero-Noise Extrapolation (ZNE), Probabilistic Error Cancellation (PEC), Clifford Data Regression (CDR), and Measurement Error Mitigation (MEM)~\cite{Temme2017PEC, Mari2021NEPEC, cai2023quantum, Bultrini2023unifying}.
In ZNE, the zero-noise expectation value of an operator is extrapolated from artificially noise-scaled circuits~\cite{Temme2017PEC, Kandala2019, Tiron2020DigitalZNE}.
In PEC, a target (ideal) circuit is approximated by averaging over distinct noisy circuits which consist of noisy, but implementable, quantum gates.
The expectation values of the operator for the noisy circuits are combined to approximate the zero-noise expectation value for the target circuit~\cite{Zhang2020, Mari2021NEPEC}.
CDR executes a group of near-Clifford circuits on a simulator and a quantum device, where the near-Clifford circuits are quantum circuits (collectively similar to the target circuit) composed largely of Clifford gates (gates that map Pauli operators to Pauli operators). It then
utilizes linear regression or machine-learning
methods to infer the zero-noise expectation value for the target circuit via the expectation values obtained from the near-Clifford circuits~\cite{Czarnik2021errormitigation, 2021CDR2}.
MEM aims to reduce measurement errors by generating a calibration matrix, whose inverse is utilized to compensate for the measurement errors~\cite{MeasurementQiskit}.

In this work, we introduce variants of ZNE, which, rather than focusing on extrapolation to zero-noise limits of expectation values, consider extrapolation of the probabilities of eigenvalue outcomes - the eigenvalues being those used to construct expectation values. 
From various extrapolations of these probabilities to the zero-noise limit, an estimated probability is then selected (calculated). 
A second difference, relative to standard ZNE, is that our ZNE variants are designed with a performance metric of an application in mind, rather than a focus on expectation values.
As we shall see, by considering \textit{both} these differences simultaneously, much improved performance on current ZNE algorithms can be delivered. 

We will consider the implementation of our algorithms on a current NISQ device - a superconducting quantum device manufactured by IBM~\cite{ibmq}.
This device is manipulated via IBM's open-source software development kit---the Quantum Information Science toolKit (Qiskit)~\cite{qiskit2024}.
Via Qiskit, we operate a simulator, \textit{ibmq\_qasm\_simulator}, and a 27-qubit quantum device, \textit{ibm\_cairo}.

\begin{figure*}[t]
\centering
\subfloat[]{\includegraphics[width=0.36\linewidth]{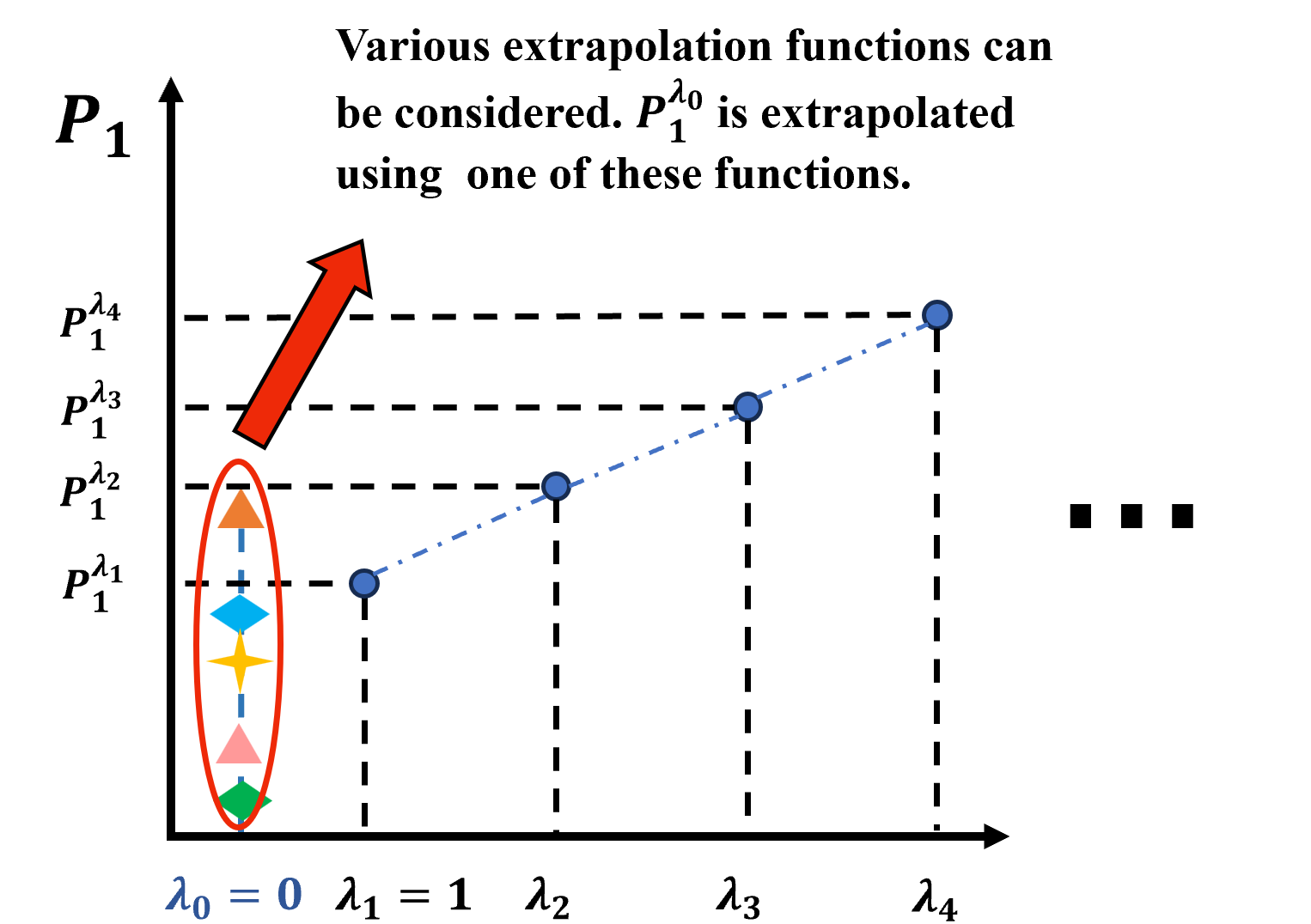}%
\label{fig:sZNEstr1}}
\hfil
\subfloat[]{\includegraphics[width=0.29\linewidth]{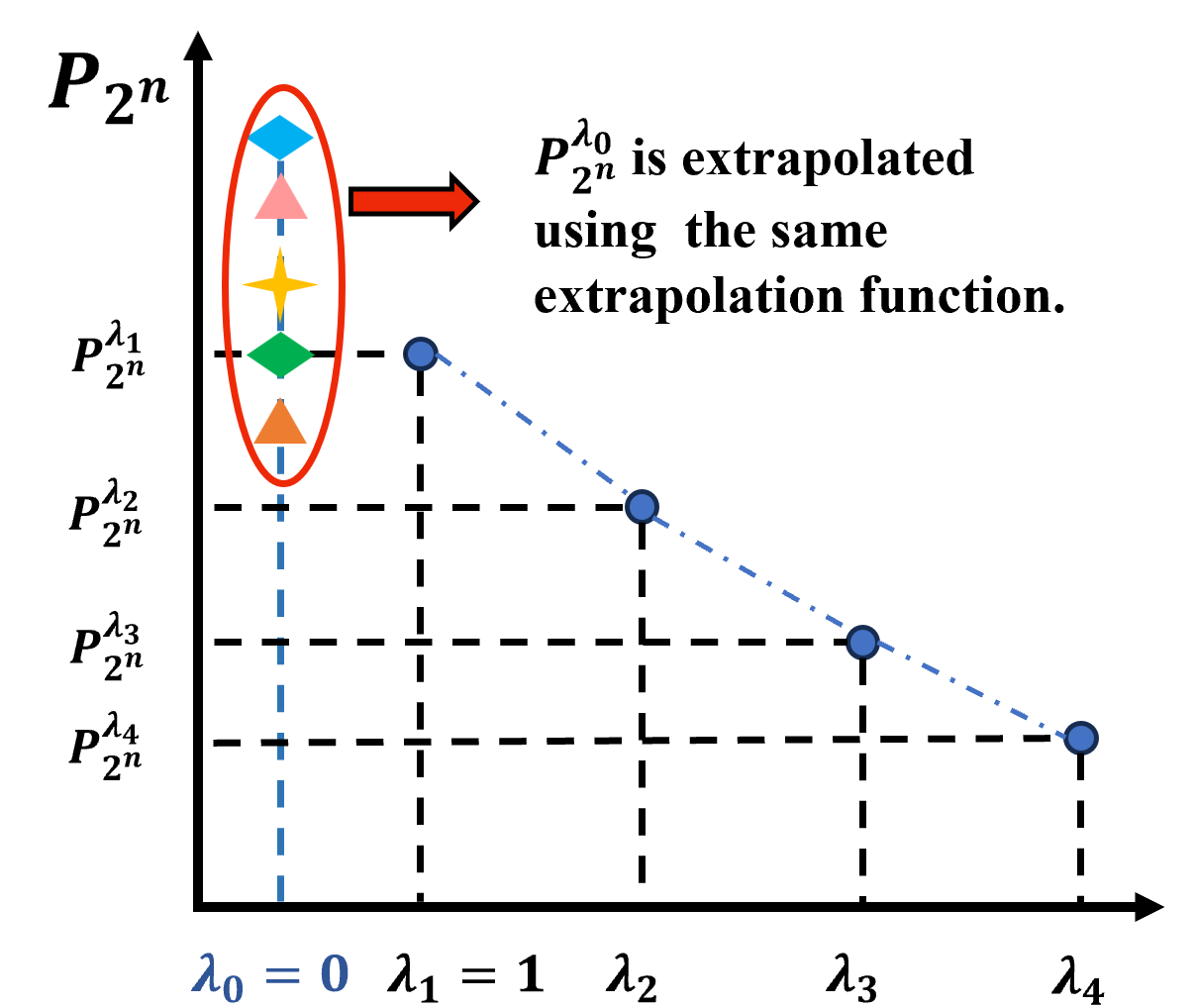}%
\label{fig:sZNEstr2}}
\hfil
\subfloat[]{\includegraphics[width=0.3\linewidth]{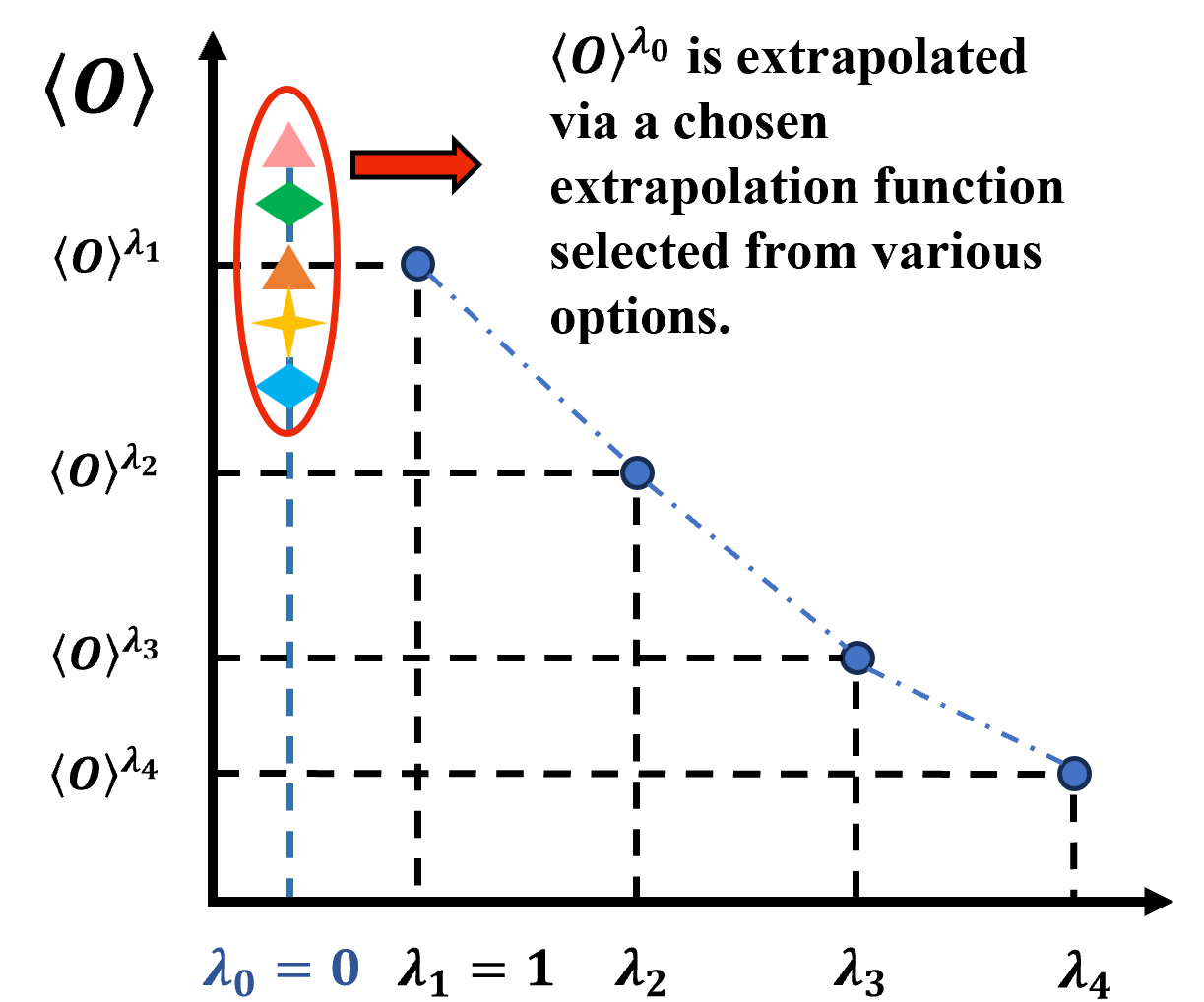} %
\label{fig:ZNEstr}}
\caption{
(a), (b) Schematic of ZNE shows how it calculates the zero-noise probabilities. 
The square dots between subfigures represent the intervening probabilities $P_{2}$ to $P_{2^n-1}$.
(c) Schematic of ZNE that calculates the zero-noise expectation value.
Note that in traditional ZNE, only one extrapolation function is selected to extrapolate the zero-noise probabilities or the zero-noise expectation value. }
\label{fig:Str}
\end{figure*}

Further,
in order to investigate the performance of our algorithms we will apply them to a specific application, namely,
Quantum Random Access Memory (QRAM)  - a system for storing both classical \emph{and} quantum information in an array of memory cells~\cite{QRAM2008, Arunachalam2015on, qramFault2020Matteo}.
QRAM can be considered a quantum version of classical RAM, with the added ability to query multiple addresses simultaneously. 
QRAM can also provide the fundamental architecture for realizing quantum oracles~\cite{Matteo2020QRAM, Paler2020QRAM, phalak2023quantum}.
We couple our modified algorithms with tomography circuits to examine the output state of QRAM implemented on the device and investigate the fidelity of the output state compared to the ideal QRAM output state.
One can envisage deployment scenarios where such checks are carried out on some test fraction of a state ensemble. 
Our algorithms apply mitigation to all aspects of this fidelity check.

The two main contributions of this work are summarized as follows.
(i)~We propose a modified version of ZNE, henceforth referred to as selected-ZNE (sZNE). 
We consider two different methods, within sZNE, of selecting the preferred extrapolation function, one based on some noisy estimate of the noiseless limit (from an independent method), and one independent of such a noisy estimate. 
The former method allows integration of sZNE with third-party algorithms, and the latter method provides a stand-alone solution. 
Both methods are detailed in Section~\ref{select}.  (ii) We then embed ZNE and sZNE into a quantum device and experimentally deploy both algorithms in the context of QRAM (with quantum state tomography) in order to show their relative real-world performance  (fidelity). This is detailed in Section~\ref{EX}.  

We commence our study with some required background material.


\section{Background} \label{Background}

Before proceeding let us clarify some of the metrics and variables we will use in our analysis. 
Suppose that $O$ is an operator with expectation value $\left< O \right>$ and discrete eigenvalues $\{a_x, \,  x = 1,2, \cdots, 2^n\}$, each associated with one of the $2^n$ eigenvectors.
Here, $n$ is the number of some qubits (represented by a state $\vert \psi \rangle$) to be measured by the operator $O$.
In the operational perspective, $C_{tot}$ copies of $\vert \psi \rangle$ are prepared, and each copy is measured by $O$.
The number of times that $a_x$ is obtained as the result of the measurement is denoted as $C_x\in [0, C_{tot}]$.
The frequency of obtaining $a_x$ from a limited number of trials is defined as $F_x = C_x/ C_{tot}$. 
The  probability  of obtaining $a_x$ with zero finite sampling error is obviously $P_x =\lim_{C_{tot} \to \infty} \left( C_x/C_{tot}\right)$. 
Therefore, the expectation value of $O$ with respect to the state $\vert \psi \rangle$ can be expressed as
\begin{equation}\label{eq:exptV1}
\left< O \right> = 
\lim_{C_{tot} \to \infty} \sum_{x=1}^{2^n} \dfrac{C_x}{C_{tot}} a_x
 = \sum_{x=1}^{2^n} P_x a_x
\text{.}
\end{equation}
Henceforth, we assume $ P_x = F_x$.

  \subsection{ZNE}

ZNE is a quantum error mitigation method that involves running additional quantum circuits and classical post-processing of experimental data.
The main idea of ZNE is to extrapolate the zero-noise expectation value of an operator from noise-scaled circuits at different noise levels~\cite{tran2023locality}.
ZNE can be divided into analog and digital ZNE based on the noise-scaling method adopted.
Analog ZNE scales the noise by extending the microwave pulse duration (used to execute a gate), while digital ZNE scales the noise via the insertion of additional quantum gates.
In this work, we consider only digital ZNE.

In digital ZNE, global and local folding are the two main methods utilized for generating noise-scaled circuits.
Global folding replaces a unitary circuit $U$ with $U\rightarrow U\left(U^\dag U\right)^{\xi}$, where $\xi$ is a positive integer.
As $U^\dag U = I$, this folding operation amplifies the noise of noisy quantum devices without adding any logical effects (see~\cite{RAM} for additional noise contributions associated with this concept).
Local folding folds a \emph{subset} of gates in $U$, where the gates in the subset are selected randomly, and each gate in the subset is folded with the same logic as global folding.
Suppose that noise-scaling factors $\boldsymbol{\lambda} = \left[\lambda_1, \lambda_2, \cdots, \lambda_j, \cdots, \lambda_J \right]$ represent the amplification of the noise in the noise-scaled circuits.
Specifically, $\lambda_j$ represents the ratio of the unitary gates in the $j^{\text{th}}$ noise-scaled circuit to the number of the unitary gates in $U$.
The values of $\boldsymbol{\lambda}$ are real numbers in ascending sequence with (typically) $\lambda_1 = 1$, and there are $J$ noise-scaled circuits in total.

Suppose that the measurement of $O$ is applied to $n$~qubits of the generated noise-scaled circuits.
After the execution of the noise-scaled circuits on the quantum device, the measurement results of these circuits are collected.
The measurement results of the noise-scaled circuit are probabilities
$\boldsymbol{P_{s}^{\lambda}} = [\boldsymbol{P_{s}^{\lambda_1}}, \cdots, \boldsymbol{P_{s}^{\lambda_j}}, \cdots, \boldsymbol{P_{s}^{\lambda_J}}]$, 
where $\boldsymbol{P_{s}^{\lambda_j}} = [P_{1}^{\lambda_j}, \cdots, P_{x}^{\lambda_j}, \cdots, P_{2^n}^{\lambda_j} ]$.
Note that $\boldsymbol{P_s^\lambda}$ can also be represented as $\boldsymbol{P_{s}^{\lambda}} = [\boldsymbol{P_{1}^{\lambda}}, \cdots, \boldsymbol{P_{x}^{\lambda}}, \cdots, \boldsymbol{P_{2^n}^{\lambda}}]$, 
where $\boldsymbol{P_{x}^{\lambda}} = [P_{x}^{\lambda_1}, \cdots, P_{x}^{\lambda_J} ]$.
We separately define the zero-noise probabilities 
$\boldsymbol{P_{s}^{\lambda_0}} = [P_{1}^{\lambda_0}, \cdots, P_{x}^{\lambda_0}, \cdots, P_{2^n}^{\lambda_0}]$, where $\lambda_0 =0 $.
For clear denotation, suppose that $\Lambda \in \{ \lambda_0, \lambda_1, \cdots, \lambda_J\}$.
The process of finding 
$\boldsymbol{P_{s}^{\lambda_0}}$ via extrapolation is illustrated in Fig.~\ref{fig:sZNEstr1} and Fig.~\ref{fig:sZNEstr2}.
Note that in the entire process of ZNE, only one extrapolation can be considered.
The least square method is typically utilized to find the best-fit parameters of a chosen extrapolation function.

Originally, ZNE was proposed with Richardson extrapolation~\cite{Temme2017PEC}, 
and this remains a commonly used extrapolation function in ongoing studies of ZNE~\cite{Mari2021NEPEC, 2021CDR2, zne2020He}.
Beyond Richardson extrapolation, other functions can be used, including linear, polynomial, poly-exponential, and exponential extrapolation~\cite{Tiron2020DigitalZNE}.
Given an extrapolation function, there are two approaches to extrapolate the zero-noise expectation value $\left< O \right>^{\lambda_0}$ using $\boldsymbol{P_{s}^{\lambda}}$.
One approach is applying the extrapolation function to noisy expectation values, denoted as $\boldsymbol{\left< O \right>^{\lambda}} = [\left< O \right>^{\lambda_1}, \cdots, \left< O \right>^{\lambda_j}, \cdots, \left< O \right>^{\lambda_J} ]$, as shown in Fig.~\ref{fig:ZNEstr}.
Note that $\left< O \right>^{\lambda_j}$ is calculated from $\boldsymbol{P_{s}^{\lambda_j}}$ using Eq.~\eqref{eq:exptV1}.
Another approach involves applying the extrapolation function to $\boldsymbol{P_{x}^{\lambda}}$ to obtain $P_{x}^{\lambda_0}$, and then calculating $\left< O \right>^{\lambda_0}$ using $\boldsymbol{P_{s}^{\lambda_0}}$ via Eq.~\eqref{eq:exptV1}.
Both approaches yield the same value for $\left< O \right>^{\lambda_0}$ since the same extrapolation function is utilized.

From the working process of ZNE, it is evident that multiple assumptions must hold true for ZNE to provide an effective error-mitigated result.
In digital ZNE, a critical assumption is that the noise in a quantum device can be amplified by folding unitary gates in $U$, implying that the noise is assumed to be incoherent errors~\cite{Tiron2020DigitalZNE}.
Incoherent errors are associated with independent gate errors and decoherence.
The execution time of the noise-scaled circuit increases with its circuit depth, resulting in greater decoherence.
Other types of errors, such as coherent errors and measurement errors, may not be amplified by global and unitary folding.
Coherent errors might be canceled by adding the inverse of a unitary gate. 
Other methods, including randomized compiling and twirling~\cite{Kurita2023synergeticquantum, twirl2022Chen}, can be considered for mitigating coherent errors.
We note measurement errors are unrelated to the circuit depth.

In ZNE, the circuit depth of the noise-scaled circuit is independent of the number of qubits in $U$.
In digital ZNE with global folding, the circuit depth of the noise-scaled circuit increases linearly with the number of gates in $U$.
In digital ZNE with local folding, the circuit depth of the noise-scaled circuit is determined by its noise-scaling factor, $\lambda_j$.
In ZNE, the best choice of the extrapolation function with its best-fit parameters is unknown \emph{a priori.}
The accuracy of the extrapolated values depends on the quantity and accuracy of the input data injected into the extrapolation function.
Generally, having more noise-scaled circuits with a wider range of circuit depths can improve the accuracy of extrapolation.
This is because a greater amount of the input data across a broader range allows for better identification of curve-fitting patterns and trends, leading to higher accuracy in selecting the extrapolation function and its parameters.
Beyond quantity, the accuracy of the input data in demonstrating the amplification of the noise is also crucial.
Moreover, extrapolation can be unreliable when the input data have critical fluctuations and/or the chosen function is a high-order polynomial extrapolation function~\cite{extrpolation}.
These problems limit the power of ZNE and lead to the fact that $\left< O \right>^{\lambda_0}$ extrapolated via ZNE can be inaccurate at times~\cite{2021CDR2}.


    \subsection{Bucket Brigade QRAM}

\begin{figure}[tb]
    \centering
    \includegraphics[width =\linewidth]{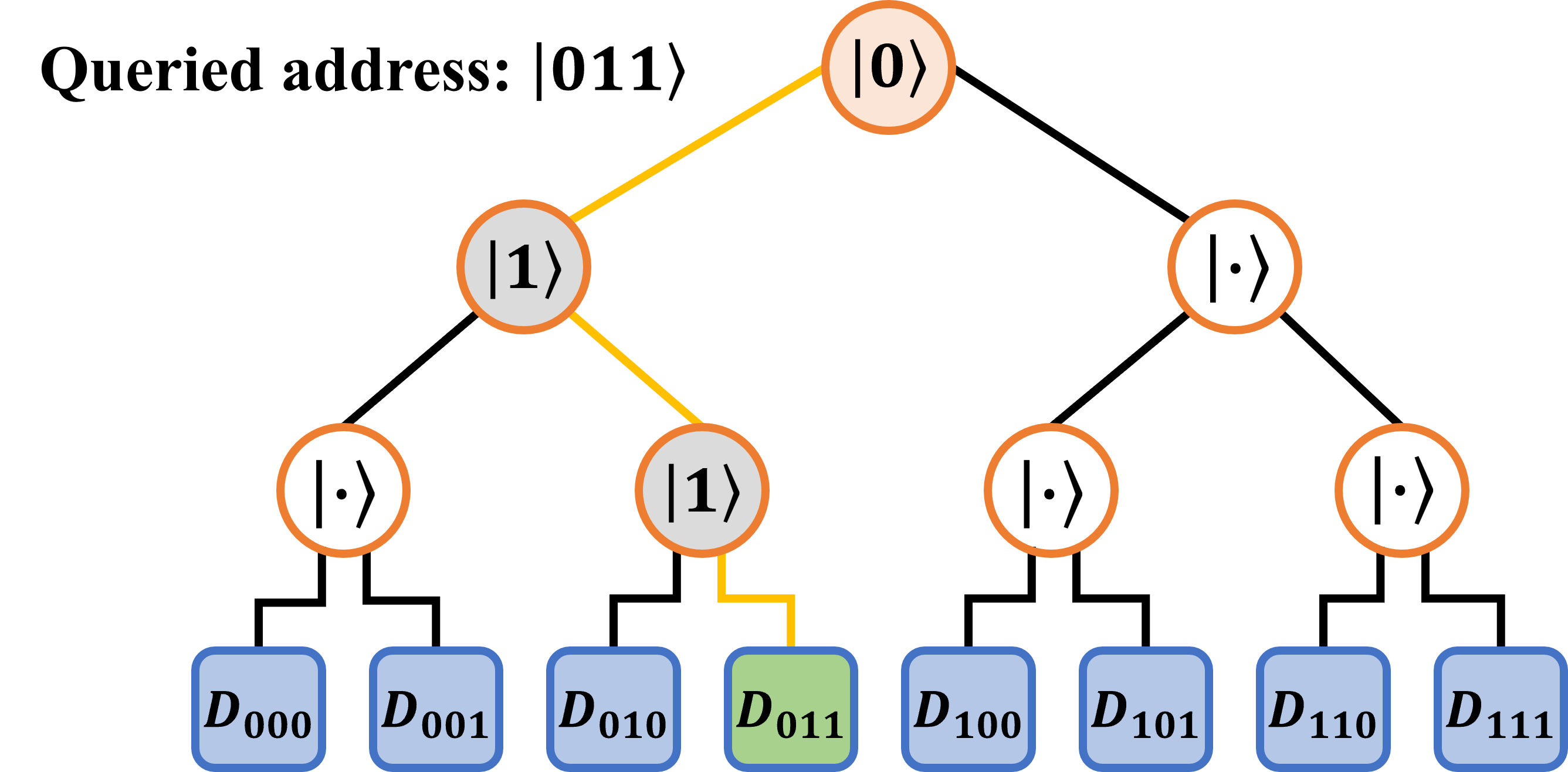}
\caption{Bucket brigade scheme for a QRAM with eight memory cells.
The binary tree nodes are initialized to the $\vert \cdot \rangle$ state, which is a waiting state.
The queried address qubits $\vert 011\rangle$ are read by the tree nodes sequentially, and the nodes that receive the address qubits will be activated and changed to the received qubits.
The activated tree nodes generate a route to the memory cell $D_{011}$.
}
    \label{fig:QRAMstr}
\end{figure}

We consider 
a tomography application that includes QRAM
to benchmark and study the performance of ZNE implemented on the quantum device.
The critical advantage of QRAM is that multiple classical and/or quantum data stored in memory cells can be queried in superposition.
A QRAM query can be expressed as
\begin{equation}
\sum_{d=0}^{N-1} \alpha_d \vert d\rangle\vert0\rangle
\xrightarrow{\text{QRAM}}
\vert\Phi\rangle_{f} = 
\sum_{d=0}^{N-1} \alpha_d\vert d\rangle \vert D_d\rangle
\text{,}
\end{equation}
where $N$ is the number of the memory cells, $\alpha_d$ is the amplitude of each address $\vert d \rangle$ in the superposition, and $\vert D_d\rangle$ represents the data stored in the memory cell addressed $\vert d \rangle$. 
One of the most robust designs for QRAM is the bucket brigade scheme, which requires a binary tree for querying addresses~\cite{Arunachalam2015on}.
A bucket brigade QRAM scheme with eight memory cells is schematically depicted in Fig.~\ref{fig:QRAMstr}.
In this QRAM, the binary tree has seven tree nodes, each of which makes a binary decision.
Each binary tree node is a three-level system (a qutrit) with states $\vert \cdot \rangle$, $\vert 0 \rangle$, and $\vert 1 \rangle$, where the $\vert \cdot \rangle$ state is called the waiting state.
In this three-level system, each tree node is initialized to the waiting state.
The queried address qubits are injected into the tree nodes from the root node, and the tree nodes read the address qubits sequentially to generate routes to the queried memory cells.
If the tree node receives the $\vert 0\rangle$ state (the $\vert 1\rangle$ state), this node activates the left (right) child, which can be either one of the next-level tree nodes or one of the memory cells.
If the queried address is $\vert 011\rangle$, the root node reads the first address qubit, which is in the state $\vert 0 \rangle$.
Then, the chosen second and third-level nodes read the rest of the two qubits sequentially.
After this reading process, the data stored in the memory cell, labeled as $D_{011}$, is accessed.

\begin{figure}[tb]
    \centering
    \includegraphics[width =0.95\linewidth]{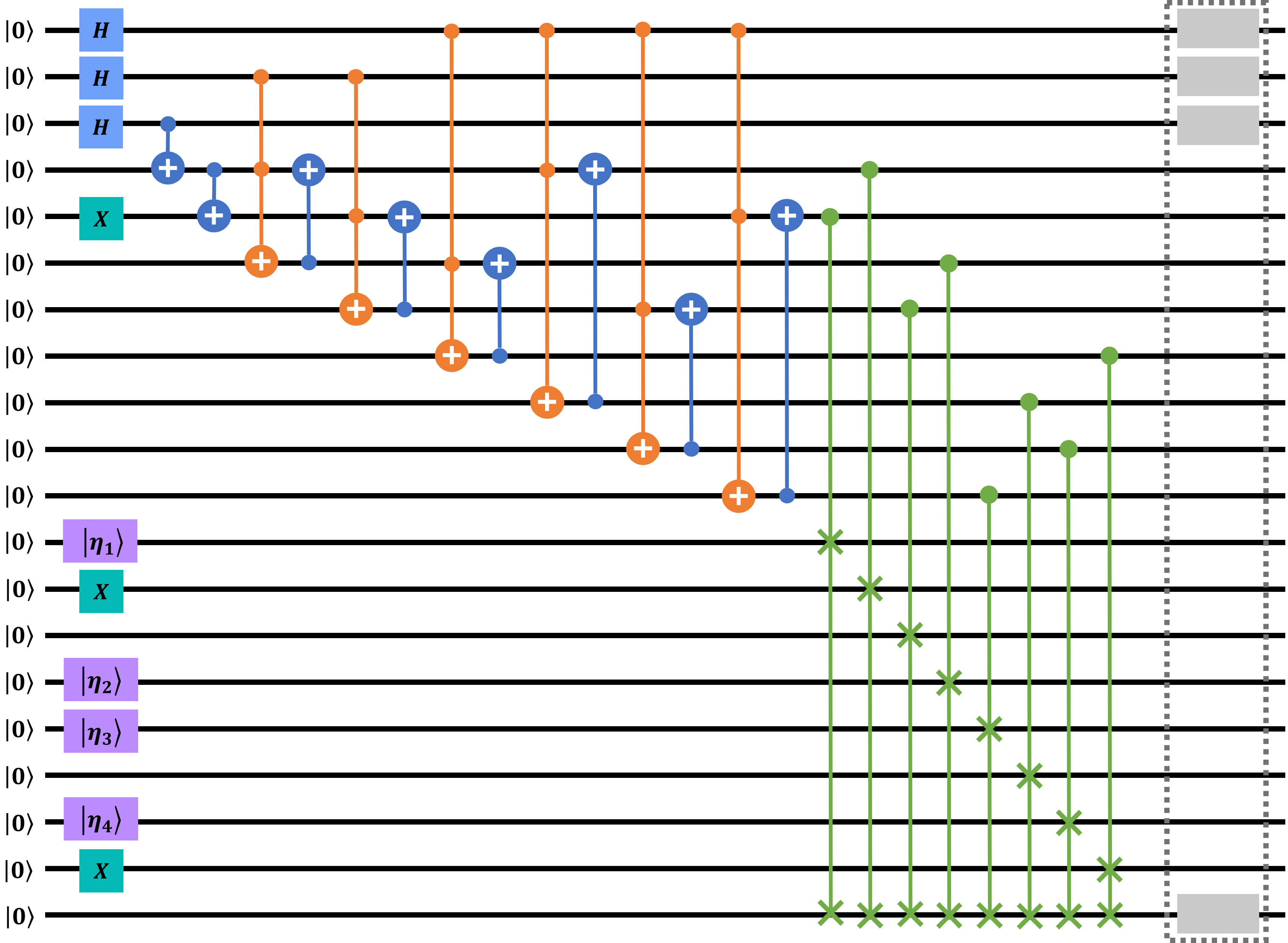}
\caption{Quantum circuit of a bucket-brigade style QRAM with quantum state tomography.
The four purple gates indicate state preparations of four random quantum states, namely, $\vert \eta_1 \rangle$, $\vert \eta_2 \rangle$, $\vert \eta_3 \rangle$, and $\vert \eta_4 \rangle$.
The $H$ represents the Hadamard gate, and the $X$ is the NOT gate.
The two-qubit gates in blue stand for $CX$ gates.
The three-qubit gates in orange and green are $CCX$ and controlled-swap gates, respectively.
The four gray boxes with the dashed line indicate that quantum state tomography is applied to these four qubits.
}
    \label{fig:QRAMcir}
\end{figure}

Based on this QRAM scheme, we construct a bucket-brigade style QRAM circuit (with quantum state tomography added), as illustrated in Fig.~\ref{fig:QRAMcir}.
The entire circuit includes 20 qubits initialized to the $\vert 0 \rangle$ state, including three address qubits, eight tree-node qubits, eight memory qubits, and one output qubit.
The first three qubits (read from the top) are the address qubits that contain the addresses intended to be queried.
Each address qubit is prepared to the state $\vert +\rangle = \left( \vert 0 \rangle + \vert 1\rangle \right) / \sqrt{2}$ via a Hadamard gate, indicating the queried addresses are $\vert 000\rangle$, $\vert 001\rangle$, $\cdots$, and $\vert 111\rangle$, \textit{i.e.}, $\sum_{d=0}^{N-1} \alpha_d \vert d\rangle = \left(\vert 000\rangle + \cdots + \vert 111\rangle \right)/ 2\sqrt{2}$. 
The next eight qubits 
are the tree-node qubits, where the second tree-node qubit is prepared to the $\vert 1\rangle$ state, and the rest to the $\vert 0 \rangle$ state.
Since qutrits cannot be implemented on the IBM quantum device, we utilize eight qubits instead of seven qutrits to act as the seven tree nodes (of Fig.~\ref{fig:QRAMstr}).
The next eight qubits of Fig.~\ref{fig:QRAMcir}
represent the memory cells (corresponding to the memory cells $D_{000}$ to $D_{111}$ of Fig.~\ref{fig:QRAMstr}), each containing a quantum state or a classical state.
The last qubit is an output qubit containing all of the queried data after the QRAM query.

Typically, once the output qubit acquires all of the queried data, the states of the binary tree nodes will be reversed to their initial state.
This step can be done by implementing the Controlled-NOT ($CX$) and Controlled-Controlled-NOT ($CCX$) gates in the QRAM circuit in reverse order, or by re-setting each tree node to the $\vert 0\rangle$ state and implementing the $X$ gate to the second tree-node qubit.
For simplicity, we ignore this simple step in the QRAM circuit.
To benchmark the performance of the QRAM,
we apply quantum state tomography to the four ``tomography qubits" - the three address qubits and the output qubit - to reconstruct $\vert\Phi\rangle_{f}$.
The quantum state tomography is represented by the gray boxes in Fig.~\ref{fig:QRAMcir}, and the details on the state tomography implementation are provided in Appendix~\ref{Apdx:QST}.
In our experiments, this QRAM tomography application is employed with ZNE or sZNE embedded into it.

\section{selected-ZNE (sZNE)} \label{select}

To benchmark the performance of ZNE embedded into the bucket brigade QRAM, we select the entanglement fidelity $F$, as our main performance metric, where $F$ is given by
\begin{equation} \label{eq:F}
F= \left( \text{Tr} \sqrt{ \sqrt{\rho} \,\, \rho' \sqrt{\rho}} \right) ^2
\text{.}
\end{equation}
Note that $\rho = \vert\Phi\rangle_{f} \langle\Phi \vert$ is the noiseless density matrix, and $\rho'$ is the density matrix that is reconstructed by quantum state tomography.
In our experiments, $U$ is the QRAM circuit which excludes the quantum state tomography.
For the execution of the state tomography (see Appendix~\ref{Apdx:QST}), 81 tomography circuits with distinct measurement operators are generated.
These measurement operators are $\{ O_g \}_{g =1}^{81} = \{ O_1 = X\otimes X\otimes X\otimes X, O_2 = X\otimes X\otimes X\otimes Y, \cdots, O_{81} = Z\otimes Z\otimes Z\otimes Z\}$, and each $O_g$ has $16$ eigenvectors (\textit{i.e.}, 16 $P_x$) as four qubits are measured.
Transpilation is necessary to execute the tomography circuits on the quantum device (see details in Section~\ref{select}).
The 81 tomography circuits are then transformed to 81 transpiled tomography circuits ($\mathbb{U}_1$ to $\mathbb{U}_{81}$).


In contrast to ZNE, which applies a chosen extrapolation function to noisy expectation values, the modified version we introduce here, sZNE, considers multiple extrapolation functions to obtain zero-noise probabilities of a measurement producing a specific eigenvalue, $a_x$, of a quantum operator possessing $2^n$ eigenvectors.
The main steps of sZNE are as follows.

\noindent (i)~Generate $J$ noise-scaled circuits with $\boldsymbol{\lambda}$, based on $\mathbb{U}_1$ (for example) using global or local folding.

\noindent (ii)~Execute the $J$ noise-scaled circuits on the quantum device and collect their measurement results, which are $\boldsymbol{P_s^\lambda}$.

\noindent (iii)~Choose an extrapolation function for each $\boldsymbol{P_{x}^{\lambda}}$ and perform the extrapolation accordingly, as shown in Fig.~\ref{fig:sZNEstr1} and Fig.~\ref{fig:sZNEstr2}.
Note that the extrapolation function selected for $\boldsymbol{P_{1}^{\lambda}}$ and $\boldsymbol{P_{2}^{\lambda}}$ (for example) can be different.

\noindent (iv) From the different extrapolated probabilities, determine a final zero-noise probability for each eigenvector.

\noindent (v)~Using the calculated zero-noise probabilities,  $\boldsymbol{P_{s}^{\lambda_0}}$, compute a performance metric of choice suited to the application under study.

As already indicated, for step (v) our application will be QRAM and our metric will be fidelity (other applications and metrics could also be studied).
In principle, an infinite number of extrapolation functions can be used, but in practice, only a limited number can be accessed.
In our experiments, we attempt Richardson extrapolation, linear extrapolation, and polynomial extrapolation with orders two and three.
Richardson extrapolation is a special case of polynomial extrapolation with order $J-1$~\cite{Tiron2020DigitalZNE}.

The fundamental idea of sZNE involves exploring various extrapolation functions for each $\boldsymbol{P_{x}^{\lambda}}$, with different functions selected to calculate them.
In contrast, ZNE exclusively employs the same extrapolation function for each $\boldsymbol{P_{x}^{\lambda}}$.
However, using the same extrapolation function may result in the introduction of additional errors.
Addressing this limitation of ZNE is the main aim of the proposed sZNE.
Since the expectation value is computed based on $P_x$, as shown in Eq.~\eqref{eq:exptV1},
it is intuitive that minimizing errors in each $P_x$  should improve any performance metric of an application that involves the use of these probabilities.
Defining that the error between $P_x^{\lambda_0}$ and noiseless simulation results $P_x^{sim}$ as 
$e_x = \left| P_x^{\lambda_0} - P_x^{sim} \right|$, 
we seek to understand how application performance metrics scale with $e_x$.


The remaining task is to identify the method utilized to determine the zero-noise probabilities in step (iv). 
There are many possibilities for this. 
Here, we focus on two different methods: 
the first method involves selecting each zero-noise probability based on a noisy estimate of the noiseless probability 
(termed the noisy estimator algorithm), and the second method uses a filter function to eliminate over-fitted solutions and calculate the zero-noise probabilities based on the solutions that pass the filter (termed the filter function algorithm).
We describe these two methods in more detail.

\subsubsection{Noisy estimator algorithm} Consider the availability of a noisy estimate, denoted as $P_{g,x}^{est}$, of the noiseless probability associated with measurement operator $O_g$ and eigenvalue $a_x$. 
This estimate could arise from several means, sources and techniques unrelated to ZNE - for our purpose it does not matter. 
We simply assume its availability and adopt no knowledge of its reliability (error). 
This represents a generic method of encapsulating a solution from a technique independent of ZNE into a new solution partially based on ZNE. 
For each $O_g$, we propose to select the zero-noise extrapolated probability for each $P_x$ as given by
\begin{equation} \label{eq:pxopt}
P_{g,x}^{\text{sZNE}} = \arg\min_{P \in L} \left| P - P_{g,x}^{est} \right|
\text{,}
\end{equation}
where $P_{g,x}^{\text{sZNE}}$ is the extrapolated probability with the selected extrapolation function, and $L = \{ P_{g,x}^{\lambda_1}, P_{g,x}^{\lambda_0, f_1}, P_{g,x}^{\lambda_0, f_2}, P_{g,x}^{\lambda_0, f_3}, P_{g,x}^{\lambda_0, f_4}\}$, where $f_1$ to $f_4$ indicate linear, polynomial extrapolation of orders 2 and 3, and Richardson extrapolation functions, respectively.
Note that for each $O_g$, the values in the set $\{P_{g,x}^{\lambda_0, f_1} \}_{x=1}^{16}$ are normalized (the same applies to other sets with different extrapolation functions), and the values in the set $\{P_{g,x}^{\text{sZNE}}\}_{x=1}^{16}$ are normalized again before reconstructing $\rho'$.
We further refer to the method with the case where $L$ excludes $P_{g,x}^{\lambda_1}$ as $\rm{sZNE}'$ to distinguish it from sZNE.
Various solutions, including Clifford simulators and the CDR method, can be considered to determine $P_x^{est}$, as discussed in detail in Appendix~\ref{Apdx:Pest}.
However, for simplicity, we generate this estimate by introducing Gaussian noise with variance $\sigma^2$ to the noiseless simulation results, $P_{g,x}^{sim}$.
That is, 
\begin{equation}
P_{g,x}^{est} = P_{g,x}^{sim} + \epsilon
\text{,}
\end{equation}
where $\epsilon$ is a random variable given by a zero-mean Gaussian distribution, \textit{i.e.},  $\epsilon \sim \mathcal{N}(0,\,\sigma^{2})$.


\subsubsection{Filter function algorithm}
For each $P_x$ in every $O_g$, we obtain the set $T_{g,x} = \{P_{g,x}^{\lambda_0, f_1}, P_{g,x}^{\lambda_0, f_2}, P_{g,x}^{\lambda_0, f_3}, P_{g,x}^{\lambda_0, f_4} \}$.
There are a total of $81\times 16$ $T_{g,x}$, and the values in each $T_{g,x}$ are passed through a filter function.
The filter function has following requirements:
(i)~We delete any elements (extrapolated probabilities) in $T_{g,x}$ that are smaller than zero since probabilities cannot be negative.
(ii)~If $P_{g,x}^{\lambda_1} \geq P_{g,x}^{\lambda_J}$, then we delete the elements in $T_{g,x}$ that are smaller than $P_{x}^{\lambda_1}$.
(iii)~If $P_{g,x}^{\lambda_1} < P_{g,x}^{\lambda_J}$, then we delete the elements in $T_{g,x}$ that are larger than $P_{g,x}^{\lambda_1}$.
After the filtering, we store the remaining extrapolated probabilities in the set $T'_{g,x}$.
Finally, we calculate the zero-noise extrapolated probabilities following
\begin{equation} \label{eq:filter}
P_{g,x}^{\text{filter}} = \left( \max L' + \min L' \right)/2
\text{.}
\end{equation}
Note that the value of $P_{g,x}^{\text{filter}}$ is obtained by averaging the maximum and minimum values in the set $L'$, where $L' = \{P_{g,x}^{\lambda_1}, T'_{g,x} \}$.
After obtaining $P_{g,x}^{\text{filter}}$ for each $x$, the values in the set $\{P_{g,x}^{\text{filter}}\}_{x=1}^{16}$ are normalized.
Finally, we use the $81\times16$ normalized zero-noise extrapolated probabilities to reconstruct $\rho'$ and then calculate $F$ (see Appendix~\ref{Apdx:QST}).

\begin{table}[t]
\caption{Detailed Information of the Memory Data}
\label{tab:eta}
    \centering
    \begin{tabular}{ccc}
    \toprule
     Quantum State    & $\alpha$ & $\beta$\\
\midrule
    $\vert \eta_1\rangle$  & $0.82+0.26i$ & $0.43-0.28i$\\
\midrule
    $\vert \eta_2\rangle$ & $-0.62+0.51i$ & $-0.15-0.57i$ \\
\midrule
    $\vert \eta_3\rangle$ & $0.25+0.74i$ & $0.31+0.54i$ \\
\midrule
    $\vert \eta_4\rangle$ & $0.44+0.56i$ & $-0.59-0.38i$ \\
\bottomrule
    \end{tabular}
\end{table}

    \section{Experimental Results} \label{EX}


We embed ZNE and sZNE into the tomography application containing QRAM to investigate their performance when implemented on \textit{ibm\_cairo}.
Transpilation is necessary for executing any quantum circuit on this device.
This process converts quantum gates to basis quantum gates,
enabling them to be executed physically and directly on the device.
The basis gates are $CX$, $I$, $X$, $\sqrt{X}$, and $RZ$ gates, where the $RZ$ gate rotates a single-qubit about the $Z$-axis with a phase factor.
Therefore, the $81$ tomography circuits need to be transpiled to the transpiled tomography circuits first.
In our experiments, $\boldsymbol{\lambda} = [1, 1.4, 1.7, 2.1, 2.5]$ and we fold gates locally at random to generate the noise-scaled circuits, 
where each transpiled tomography circuit is converted to five transpiled noise-scaled tomography circuits.
The transpilation process is realized via Qiskit, and Mitiq~\cite{LaRose2022mitiqsoftware} is utilized to generate the noise-scaled circuits and to implement extrapolation functions.
Each transpiled noise-scaled tomography circuit is executed $C_{tot} = 10,000$ times on the quantum device.

Note, the  eight qubits corresponding to the memory cells $D_{000}$ to $D_{111}$ of Fig.~\ref{fig:QRAMstr}, must be initialized.
In the results shown here we have adopted a memory allocation given by
$\vert \eta_1\rangle$, $\vert 1\rangle$, $\vert 0 \rangle$, $\vert \eta_2\rangle$, $\vert \eta_3 \rangle$, $\vert 0 \rangle$, $\vert \eta_4 \rangle$, and $\vert 1\rangle$ (as shown in Fig.~\ref{fig:QRAMcir}), where  $\vert \eta_1\rangle$ to $\vert \eta_4\rangle$ are four random quantum states. 
Each one of these four quantum states can be represented by $\alpha \vert 0 \rangle + \beta \vert 1 \rangle$:
the selected $\alpha$ and $\beta$ of the four quantum states stored in the memory cells are given in Table.~\ref{tab:eta}.

\begin{figure}[tb]
\centering
\includegraphics[width=\linewidth]{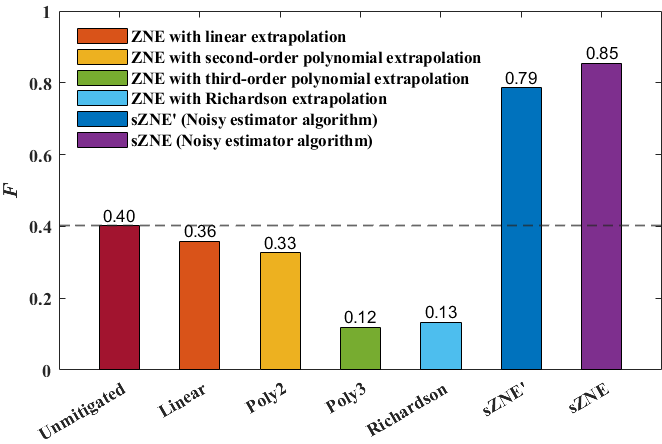}
\caption{$F$ of the QRAM with eight memory cells with or without quantum error mitigation obtained from \textit{ibmq\_cairo}. 
The $\rm{sZNE}'$ and sZNE results are calculated with $\epsilon = 0$ and in effect only show the importance of selecting the correct extrapolation function - in reality, the performance shown cannot be achieved since $\epsilon$ is always non-zero.
Note that the horizontal dashed line indicates the unmitigated $F$ for reference.}
\label{fig:FbarQRAM8m}
\end{figure}

\begin{figure}[tb]
    \centering
    \includegraphics[width =\linewidth]{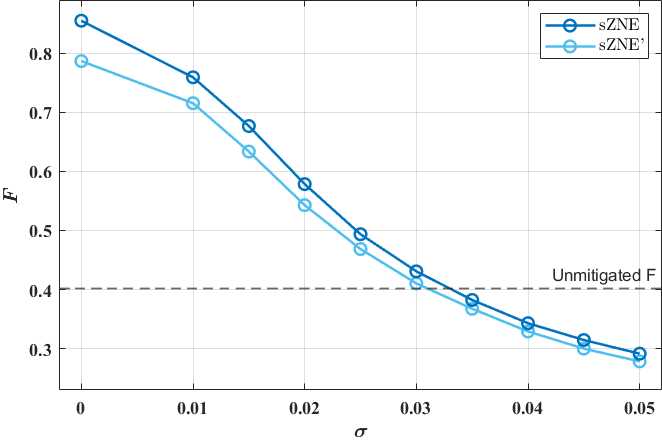}
\caption{
$F$ of the QRAM with eight memory cells as a function of the standard deviation, $\sigma$, of the Gaussian noise for $\rm{sZNE}'$ and sZNE. 
These results represent a more realistic scenario with $\epsilon \ne 0$.
The dashed line shows the unmitigated results. 
}
    \label{fig:F_Gaussian}
\end{figure}

\subsubsection{Noisy estimator results}

We consider the results from the noisy estimator algorithm. 
We first consider the $\epsilon = 0$ case in sZNE. 
Clearly, this case holds no value for mitigation (if we knew the zero-noise probability exactly, there is no need for mitigation).  
However, we use it here merely to evaluate the significance of employing proper extrapolation functions to the noisy probabilities.
The fidelity results of the tomography application that includes the QRAM with eight memory cells are illustrated in Fig.~\ref{fig:FbarQRAM8m}, and we see that $\rm{sZNE}'$ and sZNE provide for enhancement in $F$, it being increased from $0.4$ to $0.79$ and $0.85$, respectively.
The fidelity result of sZNE indicates the optimal performance of the QRAM tomography application that ZNE can achieve if a proper extrapolation function is applied to each $P_x$.

We next conduct $\rm{sZNE}'$ and sZNE with $P_{g,x}^{est}$, which is generated by introducing the Gaussian noise to the noiseless simulation results. 
This represent a more real-world scenario in which an independent estimate of the noiseless probability value is made available. 
We wish to explore what constraints must be imposed on the noisy estimation in order for our mitigation method to offer advantages in QRAM fidelity.
The results are shown in Fig.~\ref{fig:F_Gaussian},
where each circle represents a fidelity result averaged from $1,000$ repetitions.
We see that sZNE provides improved fidelity results if $\sigma$ is smaller than approximately 0.03.

\begin{table}[t]
\caption{Experimental Results of the Filter Function Algorithm}
\label{tab:filter}
    \centering
    \begin{tabular}{ccc}
    \toprule
 & Fidelity &   $\{\left< O_g \right>\}_{g=1}^{81}$ Performance   \\
\midrule
Filter Function Algorithm & 0.48 & $19\%$ \\
\bottomrule
    \end{tabular}
\end{table}

\subsubsection{Filter function results}

\begin{figure}[t]
\centering
\includegraphics[width=\linewidth]{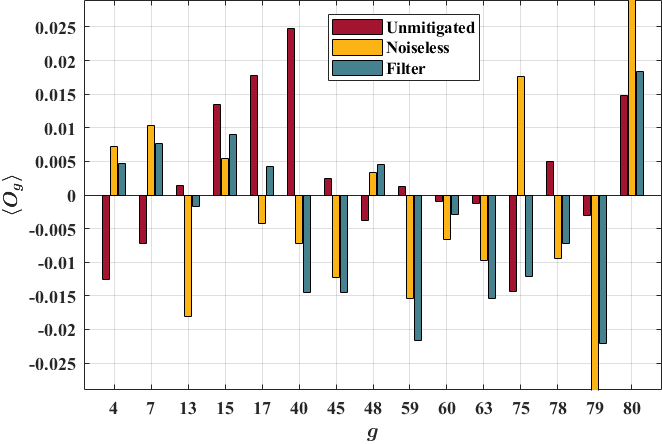}
\caption{
Expectation values of $O_g$ with and without the filter function algorithm.
The noiseless expectation values calculated via $P_{g,x}^{sim}$ are shown for reference.
Only 15 of the 81 expectation values, where the error-mitigated values (using the filter function algorithm) are closer to the corresponding noiseless ones compared to the unmitigated values, are illustrated.
}
\label{fig:filtersEV}
\end{figure}

We consider the results from the filter function algorithm, as shown in Table.~\ref{tab:filter}.
We see that this algorithm improves the fidelity result of the QRAM tomography application from 0.4 to 0.48, which is the main result of this paper.
This represents a $20\%$ improvement in the key metric of our application, and shows the merit of our approach.
Since the expectation value of an observable is the typical performance metric used in quantum error mitigation, we also demonstrate the performance of the filter function algorithm in terms of $\left< O_g \right>$, compared to the unmitigated values.
Using the filter function algorithm, we find only 15 error-mitigated expectation values (see Fig.~\ref{fig:filtersEV}) among the 81 $\left< O_g\right>$ are closer to their corresponding noiseless values  - a $19\%$ performance level.

From the above discussion, we learn that using the unmitigated $\left< O_g \right>$ directly produces better results overall in determining expectation values, relative to our new filter function algorithm. 
This is in contrast to the result achieved when looking at the performance metric of our application directly, the fidelity. 
This counter-intuitive result illustrates the advantage of our algorithm. 
Negating the traditional use of ZNE and its focus on expectation values of observables, but rather bypassing these values and focusing on the performance metric of the application instead can produce useful outcomes.

\section{Conclusions} \label{Conclu}

We introduced a modified ZNE, referred to as sZNE, and applied it to a QRAM application with quantum state tomography on a NISQ device.
Novel in our approach was a focus on application performance metrics in the design process rather than on expectation values as in standard ZNE. 
A specific new element was our direct use of extrapolated zero-noise probabilities, which are directly coupled to QRAM fidelity, thereby circumventing the need for expectation values of operators.
To implement a QRAM with multiple memory cells on a 27-qubit quantum device, we first conducted sZNE coupled to an algorithm based on a noisy estimate of the noiseless probability available from an independent technique. 
As the error in that estimate approached zero, our calculations reduced to a study in how a judicious choice of extrapolation function on a per measurement basis can dramatically improve the ZNE technique. 
Additional experimental results demonstrated the noise threshold below which this form of sZNE is effective.
We then coupled our sZNE method to an algorithm that did not require an independent noisy estimate on the noiseless probability, but rather one based on avoidance of over-fitting and subsequent use of the remaining extrapolated probabilities,  showing how significant fidelity gain in our QRAM application can be found.

In the NISQ era, quantum error mitigation is likely to remain a critical method for near-term quantum applications.
Improving and benchmarking error mitigation methods will remain important tasks and will ensure  near-term quantum devices continue to play a constructive role in the path towards full fault-tolerant quantum computing. 
We believe the results shown here illustrate that deviations from established pathways for error mitigation still hold promise in this regard, especially if the focus is on the application metrics of interest.


\section*{Acknowledgment}
We acknowledge the use of IBM~Quantum services for this work and the advanced services provided by IBM~Quantum Hub at the University of Melbourne. RM acknowledges insightful discussions with Martin Savage on a visit to the University of Washington.
WS is supported by the China Scholarship Council, the University of New South Wales, and the Sydney Quantum Academy, Sydney, NSW, Australia.
NKK acknowledges the support from the INSPIRE Faculty Fellowship awarded by the Department of Science and Technology, Government of India (Reg. No.: IFA22‐ENG 344) and the New Faculty Seed Grant from the Indian Institute of Technology Delhi.
MRM is supported by an Australian Research Council Future Fellowship, funded by the Australian Government (project number FT200100928).


\appendices
\section{Quantum State Tomography} \label{Apdx:QST}


We realize quantum state tomography via Qiskit.
In doing this for our experiments, the 4-qubit density matrix $\rho'$ is reconstructed. This  can be expressed as
\begin{equation} \label{eq:qstN}
\begin{split}
\rho' &= \frac{1}{16}
\sum_{l_1,l_2, l_3,l_4 = 0}^{3} S_{l_1,l_2, l_3,l_4} \left(\hat{\sigma}_{l_1}\otimes \hat{\sigma}_{l_2} \otimes \hat{\sigma}_{l_3} \otimes \hat{\sigma}_{l_4} \right)\\
& =
\frac{1}{16} \left[ 
S_{0,0, 0,0} \left(\hat{\sigma}_{0} \otimes  \hat{\sigma}_{0} \otimes  \hat{\sigma}_{0}  \otimes \hat{\sigma}_{0} \right) 
+ \cdots \right. \\
 & \quad\quad\quad \left. +
S_{3,3,3,3} \left(\hat{\sigma}_{3} \otimes  \hat{\sigma}_{3} \otimes  \hat{\sigma}_{3}  \otimes \hat{\sigma}_{3} \right)
\right] 
\text{,}
\end{split}
\end{equation}
where $S_{l_1,l_2, l_3,l_4}$ are parameters determined by the measurement results of the tomography circuits, and $S_{0,0, \cdots, 0} = 1$ (due to normalization).
Here, $\hat{\sigma}_{0}$ to $\hat{\sigma}_{3}$ are the Pauli matrices, given by 
\begin{equation}
\begin{split}
\hat{\sigma}_{0} &= I = 
\begin{pmatrix}
1 & 0 \\ 
0 & 1
\end{pmatrix}
\text{,} \quad
\hat{\sigma}_{1} = X = 
\begin{pmatrix}
0 & 1 \\ 
1 & 0
\end{pmatrix}
\text{,} \\
\hat{\sigma}_{2} &= Y = 
\begin{pmatrix}
0 & -i \\ 
i & 0
\end{pmatrix}
\text{, and } \,
\hat{\sigma}_{3} = Z = 
\begin{pmatrix}
1 & 0 \\ 
0 & -1
\end{pmatrix}
\text{.}
\end{split}
\end{equation}

Typically, $3^n$ tomography circuits with different measurement operators are generated for a $n$-qubit state tomography.
As the state tomography is applied to the four tomography qubits, $S_{l_1,l_2, l_3,l_4}$ are determined by the outcomes of the 81 tomography circuits 
in the operational perspective.
For the tomography circuit with the measurement operator $Z \otimes X \otimes Y \otimes Z$, for example, the $Z$-, $X$-, $Y$-, and $Z$-basis measurements are applied to the four tomography qubits.
Qiskit~\cite{qiskit2024} only supports the $Z$-basis measurements, which means that to implement the $X$-basis measurement, we need to apply the Hadamard gate before the $Z$-basis measurement.
To conduct the $Y$-basis measurement, we implement the $S^{\dag}$ gate and the Hadamard gate sequentially before the $Z$-basis measurement, where $S^{\dag}$ is a phase gate which induces a $- \pi/2$ phase.

The parameters $S_{l_1,l_2, l_3,l_4}$ are determined by the results of the tomography circuits.
For instance, 
$S_{0,1,2,3} = 
\left(P_{Z_1^{+1}} + P_{Z_1^{-1}} \right) \cdot
\left(P_{X_2^{+1}} -P_{X_2^{-1}} \right) \cdot
\left(P_{Y_3^{+1}} - P_{Y_3^{-1}} \right) \cdot
\left(P_{Z_4^{+1}} - P_{Z_4^{-1}} \right) $, 
where $P_{Z_4^{+1}}$ and $P_{Z_4^{-1}}$ are the probabilities of obtaining the $+1$ and $-1$ eigenvalues, respectively, when the $Z$-basis measurement is applied to the fourth tomography qubit.
Based on the expression of $S_{0,1,2,3}$, we can find that the parameters $S_{0,1,2,3}$, $S_{0,1,2,0}$, $S_{3,1,2,0}$, and $S_{3,1,2,3}$ are all determined by the results of the tomography circuits with the measurement operator $Z \otimes X \otimes Y \otimes Z$.
After collecting the results of the $81$ tomography circuits, we can reconstruct $\rho'$ via Eq.~\eqref{eq:qstN}.

Importantly, we see now how the parameters $S_{l_1,l_2, l_3,l_4}$ required for the fidelity comparisons are determined directly from extrapolated probability results, bypassing any requirement for expectation values. 
This offers a unique use of ZNE, leading to preferred outcomes on application metrics.

\vfill

\section{Discussion on Generating $P_x^{est}$} \label{Apdx:Pest}


In the above, we have simulated $P_x^{est}$ by adding Gaussian noise to the noiseless value obtained from noiseless classical simulations. 
Other noise models could be considered, such as Gaussian noise with the addition of a bias term and non-Gaussian noise.
Clearly, if we could always classically simulate an accurate value for all probabilities we use, there would be no requirement for ZNE (or its variants).
We emphasize again that in our calculations, we have assumed $P_x^{est}$ is available from an independent technique. 
When the error in $P_x^{est}$ approaches zero, our calculation reduce to a study in how a judicious choice of extrapolation function on a per measurement basis can dramatically improve the ZNE method. 
We offer this algorithm as a means to integrate different mitigation methods with ZNE.
This is the main message of the initial part of our  study.

We have left open how to find such a reliable $P_x^{est}$ in practice from other methods.
A potential method is via the execution of a near-Clifford circuit that is approximately equal to the circuit of interest, $U$.
With fewer non-Clifford gates, the near-Clifford circuit can be executed on the simulator, even when $U$ cannot. 
The simulation results of the near-Clifford circuit will be regarded as $\{P_x^{est}\}_{x=1}^{2^n}$ and provide guidance for selecting appropriate extrapolation functions.

Another method is CDR (mentioned in the Introduction) - a learning-based error mitigation method using near-Clifford circuits to generate a linear regression model to mitigate errors - an approach that has previously delivered some useful outcomes~\cite{2021CDR2}.
The linear regression model will be applied to each $P_x^{\lambda_1}$ to obtain $P_x^{est}$.
If CDR can provide a $P_x^{est}$ in practice with reasonable accuracy (\textit{e.g.}, $P_x^{est}$ may not always allow us to select the best extrapolation function but can help eliminate the least effective ones), then the noisy estimator algorithm outlined here can provide for enhanced performance.
We suggest this approach as a possible future study not only for QRAM but also for any application running on a NISQ device.

Clearly, many other independent methods can be directly used in our algorithm. 
It is likely that, when it comes to pragmatic quantum error mitigation, no method will always prevail in providing the optimal outcome. 
Rather, a combination of different independent algorithms are likely to be of more value.

\bibliographystyle{IEEEtran}
\bibliography{IEEEabrv,References}

\end{document}